\documentclass[12pt,a4paper]{article}
\usepackage[latin1]{inputenc}   
\usepackage{latexsym}
\usepackage[dvips]{epsfig}
\title{
  {\vspace{-3cm} \normalsize \hfill
    \parbox{38mm}{MS-TPI-00-5 \\
                  cond-mat/0009016}  }\\[25mm]
Semiclassical calculation of the nucleation rate for first order
phase transitions in the 2-dimensional $\phi^4$-model beyond the
thin wall approximation
\thanks{Poster presented at the ``XIII International Congress on Mathematical
Physics'', July 17-22, 2000, London, UK.}
  }
\author{G.~M\"unster (a) and S.B.~Rutkevich (b)\\
    (a) Institut f\"ur Theoretische Physik,
        Universit\"at M\"unster,\\
        Wilhelm-Klemm-Str.~9, D-48149 M\"unster, Germany\\
        e-mail: munsteg@uni-muenster.de\\
    (b) Institute of Physics of Solids and Semiconductors,\\
        P.Brovki 17, Minsk 220072, Belarus\\
        e-mail: rut@ifttp.bas-net.by}
\date{August 1, 2000}
\begin{document}
\maketitle

\begin{abstract}
In many systems in condensed matter physics and quantum field theory,
first order phase transitions are initiated by the nucleation of bubbles
of the stable phase. Traditionally, this process is described by the
semiclassical nucleation theory developed by Langer and, in the context
of quantum field theory, by Callan and Coleman. They have shown that
the nucleation rate $\Gamma$ can be written in the form of the Arrhenius
law: $\Gamma=\mathcal{A}e^{-\mathcal{H}_{c}}$.
Here $\mathcal{H}_{c}$ is the energy of the critical bubble, and the
prefactor $\mathcal{A}$ can be expressed in terms of the determinant of
the operator of fluctuations near the critical bubble state. It is not
possible to find explicit expressions for the constants $\mathcal{A}$
and $\mathcal{H}_{c}$ in the general case of a finite difference $\eta$
between the energies of the stable and metastable vacua. For small
$\eta$, the constant $\mathcal{A}$ can be determined within the leading
approximation in $\eta$, which is an extension of the ``thin wall
approximation''. We have calculated the leading approximation of the
prefactor for the case of a model with a real-valued order parameter
field in two dimensions.
\end{abstract}
%
\section{Introduction}

The problem of the decay of the metastable false vacuum at first order phase
transitions has attracted considerable interest due to its numerous
relations with condensed matter physics \cite{RG}, quantum fields \cite{St},
cosmology \cite{Guth}, and black hole theory \cite{Kastrup}. In Langer's
theory of homogeneous nucleation \cite{Langer}, the false vacuum decay is
associated with the spontaneous nucleation of a critical bubble of a stable
phase in a metastable surrounding. In the context of quantum field theory,
the nucleation theory was developed by Voloshin et al.\ \cite{Voloshin1},
and Callan and Coleman \cite{Col,CC}. The quantity of main interest is the
nucleation rate $\Gamma$ per time and volume. In the homogeneous nucleation
theory it has the form of the Arrhenius law:
\begin{equation}
\Gamma=\mathcal{A}\exp(-\mathcal{H}_{c}), \label{Arr}
\end{equation}
where $\mathcal{H}_{c}$ is the energy of the critical bubble. The prefactor
$\mathcal{A}$ is determined by fluctuations near the critical bubble state
and can be expressed in terms of the functional determinant of the
fluctuation operator \cite{CC}.

It should be noted that in this article the so-called kinetic prefactor,
which depends on the detailed non-equilibrium dynamics of the model,
(see \cite{RG}) is not included in $\Gamma$, which is therefore equal to
twice the imaginary part of the free energy density of the metastable phase.

In the general case, it is not possible to find the explicit critical bubble
solution of the field equations analytically.  However, the problem becomes
asymptotically solvable, if the decaying metastable state is close enough in
energy to the stable one, i.e.\ if the energy density difference $\eta$
between the metastable and stable vacua is small.  The leading approximation
in this small parameter is usually called the ``thin wall approximation''
\cite{Linde}, since at $\eta \rightarrow 0$ the critical bubble radius goes
to infinity and becomes much larger than the thickness of the bubble wall.

In the thin wall approximation, the critical bubble energy $\mathcal{H}_{c}$
can be easily obtained from Langer's nucleation theory. It turns out to be
much more difficult to find explicitly the prefactor $\mathcal{A}$ in (\ref
{Arr}). This problem, which is important for applications of nucleation
theory, has been extensively studied in different models.

A remarkable result on this subject was obtained by Voloshin
\cite{Voloshin2}. He considered scalar field theory in 2 dimensions
with a potential $U(\phi)$ of the type shown in Fig.~1. Voloshin claimed
that in the limit $\eta \rightarrow 0$ the nucleation rate $\Gamma$ in
such a model can be described by the simple universal formula
\begin{equation}
\Gamma=\frac{\eta}{2\pi}\exp\left(-\mathcal{H}_{c}\right), \label{V1}
\end{equation}
where the critical bubble energy $\mathcal{H}_{c}$ is given by
\begin{equation}
\mathcal{H}_{c}=\frac{\pi\sigma^{2}}{\eta}. \label{V2}
\end{equation}
\begin{figure}[hbt]
\vspace{.8cm}
\centering
\epsfig{file=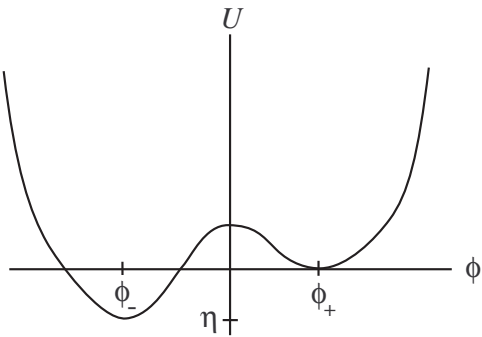,width=7cm}
\parbox[t]{0.8\textwidth}{
\caption{The potential $U$ with the false $(\phi_{+})$ and true
$(\phi_{-})$ vacuum}
}
\end{figure}
Here $\sigma$ is the surface tension of the wall between the stable and
metastable vacua in the limit $\eta \rightarrow 0.$ Thus, according to
\cite {Voloshin2}, in this limit the nucleation rate $\Gamma$ is
determined by two well defined macroscopic parameters $\eta$ and
$\sigma$. Another claim of \cite{Voloshin2} is that there are no
corrections to formula (\ref{V1}) and (\ref{V2}) proportional to powers
of the dimensionless parameter $\eta/\sigma^{2}$. Voloshin arrived at
these conclusions by an analysis performed in the thin wall
approximation. He replaced the original scalar field theory by an
effective one, which describes only fluctuations of the critical bubble
shape. This approach implies that all other fluctuations of the
original scalar field could be properly accounted for by the correct
choice of the macroscopic parameters $\eta$ and $\sigma$.

Recently an analytical method was developed \cite{MR}, which allows one to
study nucleation in the scalar field model beyond the thin wall
approximation. In \cite{MR} this method was used to calculate the nucleation
rate for the first order phase transition in the three- dimensional
Ginzburg-Landau model. In the present paper we apply the same approach to
the two-dimensional case. We calculate the nucleation rate beyond the thin
wall approximation and verify directly Voloshin's claims (\ref{V1}) and
(\ref {V2}).

Nucleation theory in two-dimensional scalar field theory has also been
studied by Kiselev and Selivanov \cite{Kiselev}, Strumia and Tetradis
\cite {ST}, and other authors. In these articles, however, different
renormalization schemes have been used and $\Gamma$ has not been
expressed in terms of macroscopic parameters $\eta$ and $\sigma$. This
makes it difficult to compare their results with Voloshin's claims.

In the article \cite{R} the nucleation rate was calculated in the
two-dimensional Ising model in a small magnetic field. If Voloshin's results
(\ref{V1}) and (\ref{V2}) are universal, they should be applicable as well
to the Ising model in the critical region. Indeed, expressions (19), (23-26)
of \cite{R} rewritten in terms of $\eta$ and $\sigma$ are in a very good
agreement with (\ref{V1}), (\ref{V2}). The exponent factors are the same,
and the prefactors differ only by the number $\pi^{2}/9\approx 1.0966$,
which is very close to unity. This small discrepancy increased our interest
in the subject of the present study.
%
%
\section{Model and notations}

We consider the two-dimensional asymmetric Ginzburg-Landau model defined
by the Hamiltonian:
\begin{equation}
\mathcal{H}(\phi)=\int\!\!d^{2}x\,\left[\frac{1}{2}
(\partial_{\mu}\,\phi(x))^{2} + U(\phi(x))\right], \label{ham}
\end{equation}
where $\phi(x)$ is the continuous one-component order parameter, and the
potential $U(\phi)$ depicted in Fig.~1 is given by
\begin{equation}
U(\phi)=U_{s}(\phi)+\frac{\eta}{2\,v}(\phi-v)+U_{0}.
\label{U}
\end{equation}
Here $U_{s}(\phi)$ denotes the symmetric part of the potential:
\begin{equation}
U_{s}(\phi)=\frac{g}{4!}\left(\phi^{2}-v^{2}\right)^{2}. \label{Us}
\end{equation}
The potential $U(\phi)$ has a metastable minimum (false vacuum) at
$\phi=\phi_{+}$ and a stable one (true vacuum) at $\phi=\phi_{-}$. The
constant term $U_{0}$ in (\ref{U}) is chosen to ensure $U(\phi_{+})=0$.

The partition function is given by the functional integral
\begin{equation}
Z=\int\!\!D\phi\,\exp\left[-\mathcal{H}(\phi)\right]. \label{Z}
\end{equation}
The inverse temperature factor has been absorbed into $\mathcal{H}$.

It is convenient to define the mass $m$ and the ``inverse temperature''
$\beta =1/T$ parameters by
\begin{equation}
m^{2}=\frac{\partial^{2}}{\partial\phi^{2}} U_{s}(\phi)\mid_{\phi=v}=
\frac{g\,v^{2}}{3},\qquad
\beta=\frac{3m^{2}}{g}, \label{m}
\end{equation}
and to introduce dimensionless variables
\begin{equation}
\tilde{x}_{\mu}=\frac{m}{2}\,x_{\mu},\quad
\tilde{\eta}=\frac{g}{2\,m^{4}}\,\eta,\quad
\varphi(\tilde{x})=\frac{\phi(x)}{v},\quad
\varphi_{\pm}=\frac{\phi_{\pm}}{v},\quad
\tilde{\mathcal{H}}=\frac{\mathcal{H}}{\beta}. \label{dim}
\end{equation}
In dimensionless variables the Hamiltonian and partition function take
the form
\begin{equation}
\tilde{\mathcal{H}}(\varphi)=\int\!\! d^{2}\tilde{x}\;
\left[\frac{1}{2}\left(\nabla\varphi\right)^{2}
+\tilde{U}\left(\varphi(\tilde{x})\right)\right] ,
\label{Ha}
\end{equation}
where
\begin{equation}
\tilde{U}(\varphi)=\frac{1}{2}\left[\left(\varphi^{2}-1\right)^{2}
-\left(\varphi_{+}^{2}-1\right)^{2}\right]+\frac{4}{3}\,\tilde{\eta}
\left(\varphi-\varphi_{+}\right), \label{Ud}
\end{equation}
and
\begin{equation}
Z=\int\!\!D\varphi(\tilde{x})\,
\exp\left[-\beta \tilde{\mathcal{H}}(\varphi)\right].
\label{ZZ}
\end{equation}
%
%
\section{The critical bubble solution}

The uniform solutions of the field equation
\begin{equation}
\delta\tilde{\mathcal{H}}/\delta\varphi(\tilde{x})=0.  \label{vac}
\end{equation}
are the stable $\varphi_{-}$ and false (metastable) $\varphi_{+}$ vacua
given by
\begin{equation}
\varphi_{\pm}=\pm 1-\frac{\tilde{\eta}}{3}\mp\frac{\tilde{\eta}^{2}}{6}
-\frac{4\tilde{\eta}^{3}}{27}+O(\tilde{\eta}^{4}). \label{fipm}
\end{equation}
The critical bubble $\varphi_{b}(\tilde{x})$ is the non-uniform radially
symmetric solution of (\ref{vac}) approaching the false vacuum at
infinity. That is,
\begin{eqnarray}
-\frac{d^{2}\varphi_{b}}{d\,\tilde{r}^{2}}
-\frac{1}{\tilde{r}}\frac{d\varphi_{b}}{d\,\tilde{r}}
+2\varphi_{b}(\varphi_{b}^{2}-1)
+\frac{4}{3}\,\tilde{\eta} &=& 0, \label{bub}\\
\lim_{\tilde{r}\rightarrow \infty}\varphi_{b}(\tilde{r}) &=&
\varphi_{+},\nonumber
\end{eqnarray}
where $\tilde{r}=\sqrt{\tilde{x}_{\mu}\,\tilde{x}_{\mu}}.$ The profile
of the critical bubble solution is shown schematically in Fig.~2. If
$\tilde{\eta}$ is small, the thin wall centered at $\tilde{r}=\tilde{R}$
divides regions of false and stable vacua outside and inside the bubble,
respectively.

\begin{figure}[hbt]
\vspace{.8cm}
\centering
\epsfig{file=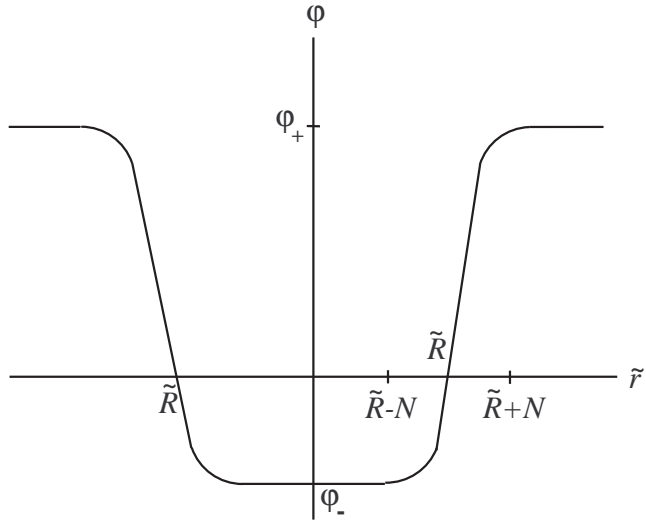,width=9cm}
\parbox[t]{0.8\textwidth}{
\caption{Profile of the critical bubble}
}
\end{figure}

Equation (\ref{bub}) can not be solved explicitly. Following the approach
introduced by M\"{u}nster and Rotsch \cite{MR} we shall construct the
solution by expansion in powers of $\tilde{\eta}$. Introducing the new
independent variable $\xi$:
\begin{equation}
\xi=\tilde{r}-\tilde{R}, \label{ksi}
\end{equation}
we write the following series for $\tilde{R}$ and $\varphi_{b}(\xi)$:
\begin{eqnarray}
\tilde{R} &=& \frac{a_{-1}}{\tilde{\eta}}+a_{0}+a_{1}\tilde{\eta}
+a_{2}\tilde{\eta}^{2}+O(\tilde{\eta}^{3}),\label{R1}\\
\varphi_{b}(\xi) &=& \varphi_{0}(\xi)+\varphi_{1}(\xi)\,\tilde{\eta}
+\varphi_{2}(\xi)\tilde{\eta}^{2}+O(\tilde{\eta}^{3}). \label{fi1}
\end{eqnarray}
After substitution of (\ref{ksi}-\ref{fi1}) into (\ref{bub}) one obtains
perturbatively in $\tilde{\eta}$:
\begin{eqnarray}
a_{-1} &=&\frac{1}{2},\;a_{0}=0,\;a_{1}=-\frac{2}{9},\;a_{2}=0,
\label{asy}\\
\varphi_{0}(\xi) &=& \tanh\xi,\;
\varphi_{1}(\xi)=-\frac{1}{3},\nonumber\\
\varphi_{2}(\xi) &=& -\frac{1}{24\cosh^{2}\xi}\,\;
\biggl\{10\xi-16\xi\cosh(2\xi)-2\xi\cosh(4\xi)+\nonumber\\
&&2\ln\left[2\cosh\xi\right]\left[\xi+8\sinh(2\xi)+\sinh (4\xi)
\right]-24\int\limits_{0}^{\xi}\!\!dt\,\,t\tanh t\biggl\}. \nonumber
\end{eqnarray}
The bubble energy
$\tilde{E}=\tilde{\mathcal{H}}\left[\varphi_{b}(x)\right]$
can be written as
\begin{equation}
\tilde{E}=\pi\int\limits_{-\tilde{R}}^{\infty}\!\!d\xi\,(\tilde{R}+\xi)
\left(\frac{d\varphi_{b}(\xi)}{d\xi}\right)^{2}. \label{ener}
\end{equation}
Substitution of (\ref{asy}) into (\ref{ener}) yields
\begin{equation}
\tilde{E}=2\pi\left[\frac{1}{3\,\tilde{\eta}}
+\tilde{\eta}\left(\frac{19}{54}-\frac{\pi^{2}}{9}\right)
+O(\tilde{\eta}^{3})\right].  \label{E}
\end{equation}
It is the basic principle of homogeneous nucleation theory that the
decay of the metastable vacuum occurs through nucleation of the critical
bubble. Callan and Coleman expressed the nucleation rate $\Gamma$ of
the metastable vacuum in terms of functional determinants \cite{Col,CC}.
In our notation their result takes the form
\begin{equation}
\tilde{\Gamma}=\frac{\beta\,\tilde{E}}{2\pi}
\frac{1}{\sqrt{\left|\lambda_{0}\right|}}
\exp\left(-\beta\,\tilde{E}+S\right). \label{W}
\end{equation}
Here $\tilde{\Gamma}=4\Gamma/m^{2}$ is the dimensionless nucleation
rate, and the entropy $S$ associated with the critical bubble is given
by
\begin{equation}
\exp S=\left[\frac{\det\mathcal{M}^{(0)}}{\det^{\prime}\mathcal{M}}
\right]^{1/2}, \label{S}
\end{equation}
where $\mathcal{M}$ and $\mathcal{M}^{(0)}$ are the fluctuation
operators near the bubble $\varphi_{b}(\tilde{x})$ and metastable
uniform vacuum $\varphi_{+}$, respectively:
\begin{eqnarray}
\mathcal{M} &=& \mathcal{-\partial}^{2}
+6\,\left[\varphi_{b}(\tilde{r})\right]^{2}-2, \label{M} \\
\mathcal{M}^{(0)} &=& \mathcal{-\partial}^{2}+6\,\varphi_{+}^{2}-2.
\label{M0}
\end{eqnarray}
The operator $\mathcal{M}$ has two zero modes proportional to
$\mathcal{\partial}_{\mu}\,\varphi_{b}(\tilde{x})$, $\mu =1,2,$ and one
negative mode with the eigenvalue
\begin{equation}
\lambda_{0}=-4\,\tilde{\eta}^{2} . \label{lam0}
\end{equation}
The notation $\det^{\prime}$ implies that the three above mentioned
modes are omitted in the corresponding determinant. After substitution
of (\ref{E}) and (\ref{lam0}), equation (\ref{W}) simplifies to
\begin{equation}
\tilde{\Gamma}=\frac{\beta}{6\,\tilde{\eta}^{2}}
\exp\left(-\,\frac{2\pi\beta}{3\,\tilde{\eta}}+S\right)
\left(1+o(\tilde{\eta}^{0})\right).
\label{Gam}
\end{equation}
In the subsequent sections we shall calculate the small $\tilde{\eta}$
expansion for the critical bubble entropy (\ref{S}) with accuracy
$o(\tilde{\eta}^{0})$.
%
%
\section{The bubble entropy}

Let us expand the bubble entropy $S$ into the sum over the angular
quantum number $\mu$:
\begin{equation}
S=\sum\limits_{\mu=-\infty}^{\infty}S_{\mu}, \label{1}
\end{equation}
where
\begin{equation}
S_{\mu}=\left\{
\begin{array}{ll}
\frac{1}{2}\sum\limits_{j=0}^{\infty}\left(\ln\lambda_{j\mu}^{(0)}
-\ln\lambda_{j\mu}\right) & \mbox{for}\ \mu\neq 0,\pm 1 \\
\frac{1}{2}\ln\lambda_{0\mu}^{(0)}
+\frac{1}{2}\sum\limits_{j=1}^{\infty}
\left(\ln\lambda_{j\mu}^{(0)}-\ln\lambda_{j\mu}\right) & \mbox{for}
\ \mu=0,\pm 1.
\end{array}
\right. \label{Smu}
\end{equation}
Here $\lambda_{j\mu}$ and $\lambda_{j\mu}^{(0)}$ are the eigenvalues of
the radial Schr\"{o}dinger operators corresponding to $\mathcal{M}$ and
$\mathcal{M}^{(0)}$:
\begin{eqnarray}
H_{\mu} &=& -\frac{d^{2}}{d\,\tilde{r}^{2}}
-\frac{1}{\tilde{r}}\frac{d}{d\,\tilde{r}}+\frac{\mu^{2}}{\tilde{r}^{2}}
+6\,\left[\varphi_{b}(\tilde{r})\right]^{2}-2, \label{1h} \\
H_{\mu}^{(0)} &=& -\frac{d^{2}}{d\,\tilde{r}^{2}}
-\frac{1}{\tilde{r}}\frac{d}{d\,\tilde{r}}+\frac{\mu^{2}}{\tilde{r}^{2}}
+6\,\varphi_{+}^{2}-2, \label{2h} \\
H_{\mu}\ \psi_{j\mu}(\tilde{r}) &=& \lambda_{j\mu}\ 
\psi_{j\mu}(\tilde{r}),\qquad 
H_{\mu}^{(0)}\chi_{j\mu}(\tilde{r})=\lambda_{j\mu}^{(0)}
\chi_{j\mu}(\tilde{r}). \nonumber
\end{eqnarray}
The eigenfunctions $\psi_{j\mu},\ \chi_{j\mu}$ are supposed to be
finite at the origin $\tilde{r}=0$ and to obey the zero boundary
condition at some large $L:\psi_{j\mu}(L)=\chi_{j\mu}(L)=0.$
%
%
\subsection{Calculation of the ``partial'' entropy $S_{\mu}$}

Let us first consider the case $\mu\neq 0,\pm 1.$ For such a $\mu$ the
operators $H_{\mu}$ and $H_{\mu}^{(0)}$ are positive. Therefore the
``partial'' entropy $S_{\mu}$ can be written in terms of the trace of
the resolvent operators:
\begin{equation}
S_{\mu}=\frac{1}{2}\int\limits_{C_{1}}\frac{d\lambda}{2\pi i}\ 
A(\lambda,\mu)\ \ln\lambda ,  \label{C1}
\end{equation}
where
\[
A(\lambda,\mu)=\mbox{Sp}\left[\left(\lambda-H_{\mu}\right)^{-1}
-\left(\lambda-H_{\mu}^{(0)}\right)^{-1}\right] .
\]
The integration path $C_{1}$ shown in Fig.~3 lies in the right
complex half-plane of $\lambda$ and goes around the spectra of both
operators $H_{\mu}$ and $H_{\mu}^{(0)}$ in the clockwise direction.  It
should be noted that the representation (\ref{C1}) is valid also for
$\mu =0, \pm 1$. In the latter case the lowest discrete mode
$\lambda_{0\mu} \leq 0$ becomes negative ($\mu = 0$) or zero ($\mu
=\pm 1$). These modes contribute neither to (\ref{Smu}) nor to (\ref{C1}).

After deformation of the path $C_{1}$ into the path $C_{2}$ which goes
around the logarithmic cut positioned at the negative real axis (see
Fig.~3), one obtains
\begin{equation}
S_{\mu}=-\frac{1}{2}\int\limits_{-\infty}^{0}\!\!d\lambda\ A(\lambda,\mu).
\label{dS}
\end{equation}
In this equation one can proceed to the thermodynamic limit
$L\rightarrow\infty$.

\begin{figure}[hbt]
\vspace{.8cm}
\centering
\epsfig{file=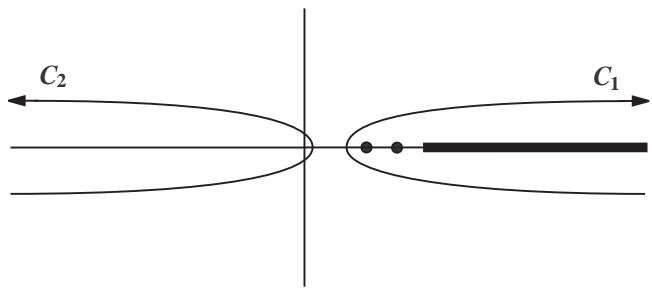,width=9cm}
\parbox[t]{0.8\textwidth}{
\caption{Integration paths $C_{1}$ and $C_{2}$ in the complex
$\lambda$-plane}
}
\end{figure}

We obtained an exact representation for the integrand in (\ref{dS}). To
describe it some notations are necessary.

Let $f_{i}(\tilde{r},\lambda),\ g_{i}(\tilde{r},\lambda),\ i=1,2$ be the
solutions of the linear ordinary differential equation
\begin{equation}
H_{\mu}\ \psi(\tilde{r})=\lambda\ \psi(\tilde{r}) \label{dif}
\end{equation}
determined by their asymptotics:
\begin{eqnarray}
f_{1}(\tilde{r},\lambda) &\rightarrow &K_{\mu}(q\tilde{r}), \ \mbox{and}\
f_{1}(\tilde{r},\lambda)\rightarrow I_{\mu}(q\tilde{r}), \ \mbox{at}\
\tilde{r}\rightarrow \infty, \label{asf} \\
g_{1}(\tilde{r},\lambda) &\rightarrow &K_{\mu}(q_{-}\,\tilde{r}),
\ \mbox{and}\ g_{1}(\tilde{r},\lambda)\rightarrow I_{\mu}(q_{-}\,\tilde{r}),
\ \mbox{at}\ \tilde{r}\rightarrow 0. \label{asg}
\end{eqnarray}
Here $K_{\mu}(z)$ and $I_{\mu}(z)$ are the modified Bessel functions, the
parameters $q$ and $q_{-}$ are defined as:
\begin{equation}
q=\left(6\varphi_{+}^{2}-\lambda-2\right)^{1/2},\quad
q_{-}=\left(6\varphi_{-}^{2}-\lambda-2\right)^{1/2}. \label{qq}
\end{equation}
Since the second order equation (\ref{dif}) has two linearly independent
solutions, there is a linear dependence between the functions
$g_{i}(\tilde{r},\lambda)$ and $f_{i}(\tilde{r},\lambda)$:
\begin{equation}
g_{i}(\tilde{r},\lambda)=\sum\limits_{j=1,2}\alpha_{ij}(\lambda)\ 
f_{j}(\tilde{r},\lambda). \label{gi}
\end{equation}
The above relation defines the $2 \times 2$-matrix $\alpha(\lambda)$.
The integrand in (\ref{dS}) can be expressed explicitly in terms of this
matrix:
\begin{equation}
A(\lambda,\mu)=\frac{3\mu(\varphi_{+}^{2}-\varphi_{-}^{2})}
{(\lambda+2-6\varphi_{-}^{2})(\lambda +2-6\varphi_{+}^{2})}
+\frac{d\ln\alpha_{22}(\lambda)}{d\lambda}, \label{tr}
\end{equation}
where $\lambda$ is assumed to be real and negative.

The representation (\ref{tr}) for the trace of resolvent operators is
exact. However, equation (\ref{dif}) can not be solved in closed form
for arbitrary $\tilde{\eta}$. So we have to consider the
small-$\tilde{\eta}$ expansion for $\alpha_{22}(\lambda)$. We have
obtained two terms of this expansion by use of a perturbation
theoretical analysis of the scattering problem (\ref{dif} -- \ref{gi}).
Omitting the details, the logarithmic derivative of the matrix element
$\alpha_{22}(\lambda)$ up to quadratic terms in $\tilde{\eta}$ takes
the form:
\begin{equation}
\frac{d\ln\alpha_{22}(\lambda)}{d\lambda} =
\frac{1}{(4+p^{2}-\lambda)^{1/2}}
\left[\frac{2\,p^{2}}{\left(\lambda-4\right)^{2}}-\frac{1}{\lambda-4}
+\frac{1}{\lambda-3-p^{2}}+\frac{2}{\lambda-p^{2}}\right]
+ O(\tilde{\eta}^{2}). \label{a22}
\end{equation}
Here $p$ is the angular momentum parameter defined as
$p=2\,\tilde{\eta}\mu\approx\tilde{R}\,\mu$, which is considered to
be a quantity of order 1 for later purposes.

Substitution of (\ref{a22}) and (\ref{fipm}) into (\ref{tr}) yields:
\begin{eqnarray}
A(\lambda,\mu) &=& \frac{1}{\left(4+p^{2}-\lambda\right)^{1/2}}
\left[\frac{2\,p^{2}}{\left(\lambda-4\right)^{2}}-\frac{1}{\lambda-4}
+\frac{1}{\lambda-3-p^{2}}+\frac{2}{\lambda-p^{2}}\right]\nonumber\\
&&-\frac{2\,\left|p\right|}{\left(\lambda-4\right)^{2}}
+O(\tilde{\eta}^{2}). \label{3}
\end{eqnarray}
Integrating (\ref{dS}) we obtain the ``partial'' entropy $S_{\mu}$:
\begin{equation}
S_{\mu}=\frac{\left|p\right|}{4}-\frac{\left(p^{2}+4\right)^{1/2}}{4}
+\frac{1}{2}\ln\left\{\frac{\left[\left(p^{2}+4\right)^{1/2}+2\right]
^{2}}{p^{2}}\,\cdot \frac{(p^{2}+4)^{1/2}+1}{(p^{2}+4)^{1/2}-1}\right\}.
\label{PS2}
\end{equation}
%
%
\subsection{Summation of the $\mu$-series}

Expression (\ref{PS2}) has a logarithmic singularity at $p=0$. This
singularity arises from the lowest discrete level of the Hamiltonian
(\ref{2h}). To account for the contribution of this level properly,
let us rewrite (\ref{1}) as
\begin{equation}
S=\sum\limits_{\mu=-\infty }^{\infty}\mbox{}s_{\mu}(z)+\ln R(z),
\label{RS}
\end{equation}
where
\begin{equation}
s_{\mu}(z)=\frac{\left|p\right|}{4}-\frac{\left(p^{2}+4\right)^{1/2}}{4}
+\frac{1}{2}\ln\left\{\frac{\left[(p^{2}+4)^{1/2}+2)\right]^{2}}
{p^{2}+z^{2}}\,\cdot\frac{(p^{2}+4)^{1/2}+1}{(p^{2}+4)^{1/2}-1}\right\},
\label{dsm}
\end{equation}
\begin{equation}
\ln R(z)=\frac{1}{2}\sum\limits_{\mu=-\infty}^{\infty}
\ln\frac{p^{2}+z^{2}}{p^{2}}, \label{za}
\end{equation}
and $z$ is some fixed positive number.

First, we shall determine $R(z).$ Taking into account the presence of
the negative and zero modes, the following corrections are necessary:
\begin{eqnarray}
\ln R(z) &=&\frac{1}{2}\sum\limits_{\mu=-\infty}^{\infty}
\ln\frac{p^{2}+z^{2}}{p^{2}}
=\frac{1}{2}\sum\limits_{\mu=-\infty}^{\infty}
\ln\frac{4\,\tilde{\eta}^{2}\mu^{2}+z^{2}}{4\,\tilde{\eta}^{2}\mu^{2}}
\longrightarrow\nonumber\\
&&\frac{1}{2}\sum\limits_{\mu=-\infty \atop \mu\neq 0,\pm1}^{\infty}
\ln\frac{4\,\,\tilde{\eta}^{2}\,\mu^{2}+z^{2}}{4\,\tilde{\eta}^{2}
(\mu^{2}-1)}+\frac{1}{2}\ln z^{2}+\ln (z^{2}+4\,\tilde{\eta}^{2}).
\label{R2}
\end{eqnarray}
These corrections arise from two facts:
\begin{enumerate} 
\item
The logarithmic singularity of (\ref{PS2}) at $p\rightarrow 0$ comes
from the first discrete level $\lambda_{0\mu}=4\tilde{\eta}^{2}(\mu^{2}-1)$
of the operator (\ref{1h}).
\item
The three eigenvalues $\lambda_{0}=-4\tilde{\eta}^{2},\ \lambda_{\pm 1}=0$
of the operator (\ref{1h}) already accounted for in (\ref{Gam}) should be
omitted in $S$ and $\ln R(z).$
\end{enumerate}

After these corrections we obtain from (\ref{R2})
\[
R(z)=z\,\left(z^{2}+4\tilde{\eta}^{2}\right)
\prod\limits_{\mu=2}^{\infty}
\frac{\mu^{2}+(z/2\tilde{\eta})^{2}}{\mu^{2}-1}
=\left(2\tilde{\eta}\right)^{3}\frac{z}{2\tilde{\eta}}
\left(1+\frac{z^{2}}{4\tilde{\eta}^{2}}\right)\ f(\frac{z}{2\tilde{\eta}}),
\]
where
\[
f(x)=\prod\limits_{\mu=2}^{\infty}
\frac{\mu^{2}+x^{2}}{\mu^{2}-1}
=\frac{2}{\pi}\frac{\sinh\left(\pi x\right)}{x(x^{2}+1)}.
\]
Therefore,
\[
R(z)=\frac{2}{\pi}\left(2\tilde{\eta}\right)^{3}
\sinh\left(\pi\frac{z}{2\tilde{\eta}}\right)
\cong\frac{1}{\pi}\left(2\tilde{\eta}\right)^{3}
\exp\left(\frac{z\pi}{2\tilde{\eta}}\right),
\]
since $\tilde{\eta}$ is small, and $z$ is a fixed number. Thus, we obtain
\begin{equation}
\ln R(z)=3\ln \left(2\tilde{\eta}\right)-\ln\pi+\frac{z\pi}{2\tilde{\eta}}.
\label{lnR}
\end{equation}
This result can also be obtained by the $\zeta$-function technique
described in \cite{MR}.

For the calculation of the first sum in the right-hand side of
(\ref{RS}) it is convenient to apply Poisson's summation formula:
\begin{equation}
\sum\limits_{\mu=-\infty}^{\infty}s_{\mu}(z)
=\sum\limits_{k=-\infty}^{\infty}\ \int\limits_{-\infty}^{\infty}\!\!d\mu\ 
s_{\mu}(z)\exp\left(2\pi i\mu k\right). \label{Po}
\end{equation}
The integrals in terms with $k\neq 0$ on the right-hand side converge.
One can easily show that these integrals (at $k\neq 0$) are of order
$\tilde{\eta}$ and can therefore be neglected in the adopted
approximation. So, it is sufficient to calculate the $k=0$ term only, which
we denote by $s(z)$.
Straightforward calculations yield:
\begin{equation}
s(z)=s(0)-\frac{\pi z}{2\,\tilde{\eta}}
=\int\limits_{-\,\infty}^{\infty}\!\!d\mu\ S_{\mu}
-\frac{\pi z}{2\,\tilde{\eta}}, \label{sm}
\end{equation}
where $S_{\mu}$ is given by (\ref{PS2}).

Now we can substitute (\ref{sm}) and (\ref{lnR}) into (\ref{RS}) and
obtain
\begin{equation}
\exp S=\frac{\left(2\,\tilde{\eta}\right)^{3}}{\pi}\exp S^{(-1)}
\left[1+o(\tilde{\eta}^{0})\right], \label{des}
\end{equation}
where
\[
S^{(-1)}\equiv\int\limits_{-\,\infty}^{\infty}\!\!d\mu\ S_{\mu}
\sim\tilde{\eta}^{-1}.
\]
As $p\rightarrow\infty,$ $S_{\mu}\sim-(2p)^{-1}.$ Therefore the
integral in $\mu$ on the right-hand side diverges logarithmically at
large $\left|\mu\right|$. Indeed, this is the well-known ultraviolet
divergency. Introducing a cut-off $p_{m}$ in $\left|p\right|$ one finds:
\begin{equation}
S^{(-1)}\leadsto\int\limits_{-\,p_{m}}^{p_{m}}\!
\frac{dp}{2\,\tilde{\eta}}\ S_{\mu}
=\frac{1}{2\,\tilde{\eta}}\left(5\ln p_{m}+\frac{11}{2}
-\frac{\pi}{\sqrt{3}}\right). \label{-1}
\end{equation}
Relations (\ref{des}), (\ref{-1}) are our final results for the bubble
entropy.
%
%
\section{Decay rate}

Substitution of (\ref{des}) into (\ref{Gam}) yields for the
dimensionless decay rate $\tilde{\Gamma}$:
\begin{equation}
\tilde{\Gamma}=\frac{4\beta\,\tilde{\eta}}{3\pi}\,
\exp\left(-\,\beta E(\tilde{\eta},T)\right),  \label{Ga}
\end{equation}
where
\begin{eqnarray}
E(\tilde{\eta},T) &=&\frac{2\pi}{3\,\tilde{\eta}}-T\ \Delta S^{(-1)}
+\,o(\tilde{\eta}^{0}) \label{E4} \\
&=&\frac{1}{\tilde{\eta}}\,\left(\frac{2\pi}{3}
-\frac{T}{2}(5\ln p_{m}+\frac{11}{2}-\frac{\pi}{\sqrt{3}})\right)
+\,o(\tilde{\eta}^{0}).\nonumber
\end{eqnarray}
To compare (\ref{Ga}), (\ref{E4}) with Voloshin's result (\ref{V1}),
(\ref {V2}), let us rewrite the latter in the dimensionless variables
(\ref{dim}) for the model (\ref{Ha})--(\ref{ZZ}):
\begin{equation}
\tilde{\Gamma}=\frac{4\beta\,\tilde{\eta}\,\langle\varphi\rangle}{3\pi}
\,\exp\left[-\,\,\frac{3\pi\,\tilde{\sigma}^{2}\,\beta}
{8\,\tilde{\eta}\,\langle\varphi\rangle}\right],  \label{ga1}
\end{equation}
where $\tilde{\sigma}$ is the surface tension, and
$\langle\varphi\rangle$ is the average value of the scalar order
parameter at $\tilde{\eta}=0.$ In (\ref{ga1}) we have replaced the
energy difference between the false and true vacua by
$\frac{8}{3}\, \tilde{\eta}\,\langle\varphi\rangle \,\beta$.

It should be stressed that the parameters $\langle\varphi\rangle$ and
$\tilde{\sigma}$ in (\ref{ga1}) are exact macroscopic quantities which
depend on the interaction parameter $g$, or in other terms, on the
``temperature'' $T\equiv\beta^{-1}=g/(3m^{2})$. At low ``temperatures''
these quantities can be expanded in $T$:
\begin{equation}
\langle\varphi\rangle=1+T\,\rho + O(T^{\,2}),\quad
\tilde{\sigma}=\frac{4}{3}+T\,\vartheta + O(T^{\,2}), \label{6}
\end{equation}
and the corrections linear in $T$ can be found perturbatively:
\begin{eqnarray}
\rho &=&-\frac{3}{4\pi}\left(\frac{1}{2}+\ln p_{m}\right),
\label{ddf} \\
\vartheta &=&-\frac{3\ln p_{m}}{\pi}-\frac{3}{\pi}+\frac{1}{2\sqrt{3}}.
\label{dds}
\end{eqnarray}
Here $\vartheta$ is the one-loop correction in $\langle\varphi\rangle$,
the correction in $\tilde{\sigma}$ was calculated in the strip geometry
from the ratio of partition functions
$\beta\tilde{\sigma}=L_{2}^{-1}\ln(Z_{0}/Z_{1})$ with different boundary
conditions:
\begin{eqnarray*}
Z_{0} &:&\quad\varphi(\tilde{x}_{1}+L_{1},\tilde{x}_{2})
=\varphi(\tilde{x}_{1},\tilde{x}_{2}+L_{2})
=\varphi (\tilde{x}_{1},\tilde{x}_{2}), \\
Z_{1} &:&\quad -\varphi(\tilde{x}_{1}+L_{1},\tilde{x}_{2})
=\varphi(\tilde{x}_{1},\tilde{x}_{2}+L_{2})
=\varphi(\tilde{x}_{1},\tilde{x}_{2}).
\end{eqnarray*}
It should be noted that both $\rho$ and $\vartheta $ contain logarithmic
ultraviolet divergencies and depend on the cut-off procedure. In
(\ref{ddf}), (\ref{dds}) we have chosen a similar cut-off-procedure as
in (\ref{-1}). Substituting (\ref{6})--(\ref{dds}) into (\ref{ga1}) and
expanding in $T$ one obtains
\[
\tilde{\Gamma}=\frac{4\beta\,\tilde{\eta}}{3\pi}\,
\exp\left[-\frac{2\pi}{3\,T\,\tilde{\eta}}+\frac{1}{2\,\tilde{\eta}}
\left(5\ln p_{m}+\frac{11}{2}-\frac{\pi}{\sqrt{3}}+O(T)\right) \right]
\]
in agreement with our results (\ref{Ga}), (\ref{E4}).
\vspace{0.3cm}

{\large \textit{\textbf{Summary}}}:
Our semiclassical calculation of the nucleation rate $\Gamma$ in the
two-dimensional Landau-Ginzburg $\phi^{4}$-model confirms Voloshin's
results (\ref{V1}), (\ref{V2}), which were derived in the thin wall
approximation. In particular, we confirm the prefactor value
$\mathcal{A}=\eta/(2\pi)$ first obtained by Kiselev and Selivanov
\cite {Kiselev}, and Voloshin \cite{Voloshin2}.

This value differs from that obtained for the two dimensional critical
Ising model \cite{R} by the numerical factor $\pi^{2}/9\approx 1.0966$.
We suppose that this small discrepancy is the result of approximations
used in \cite{R}, and the prefactor value $\mathcal{A}=\eta/(2\pi)$ is
universal.
\vspace{0.5cm}

%
\noindent
{\large \textbf{Acknowledgments}}
\vspace{0.3cm}

One of us (S.B.~R.) would like to thank the Institute of Theoretical
Physics of the University of M\"{u}nster for hospitality.
This work is supported by the Deutsche Forschungsgemeinschaft (DFG)
under grant GRK 247/2-99 and by the Fund of Fundamental Investigations
of Republic of Belarus.


%
\end{document}